\documentclass[12pt]{article}
\usepackage{graphicx}

\begin{document}

\centerline{\bf EFFECTS OF TURBULENT MIXING ON THE CRITICAL BEHAVIOR}

\bigskip

\centerline{\it N.V. Antonov, A.A. Ignatieva, A.V.Malyshev }

\

\centerline{\it Department of Theoretical Physics, St.~Petersburg State
University}

\centerline{\it Uljanovskaja 1, St.~Petersburg, Petrodvorez, 198504 Russia}

\

{\small Effects of strongly anisotropic turbulent mixing on the critical
behavior are studied by means of the renormalization group. Existence of new
nonequilibrium types of critical regimes (universality classes)
is established.}

\

\noindent{PACS 05.10.Cc, 47.27.ef}

\

Various systems of very different physical nature exhibit interesting
singular behavior in the vicinity of their critical points.
Their correlation functions reveal self-similar behavior with universal
critical dimensions: they depend only on few global characteristics of
the system (like symmetry or space dimensionality).
Quantitative description of critical behavior is provided by the
renormalization group (RG). In the RG approach, possible types of critical
regimes (universality classes) are associated with infrared (IR) attractive
fixed points of renormalizable field theoretic models. Most typical phase
transitions belong to the universality class of the
$O_{n}$-symmetric $\psi^{4}$ model of an $n$-component scalar order
parameter. Universal characteristics of the
critical behavior depend only on $n$ and the space dimensionality $d$ and
can be calculated in the form of the expansion in $\varepsilon=4-d$ or
within other systematic perturbation schemes; see the monograph \cite{Book3}
and the literature cited therein.

It has long been realized that the behavior of a real system
near its critical point is extremely sensitive to external disturbances,
geometry of the experimental setup, gravity, presence of impurities and
so on; see e.g. the monograph \cite{Ivanov} for the general discussion and
the references.
What is more, some disturbances (randomly distributed impurities or
turbulent mixing) can produce completely new types of critical
behavior with rich and rather exotic properties, like e.g. expansion in
$\sqrt{\varepsilon}$ rather than in $\varepsilon$; see e.g.~\cite{Satten}.

These issues become even more important for the nonequilibrium phase
transitions, because the ideal conditions of a ``pure'' stationary critical
state can hardly be achieved in real chemical or biological systems, and
the effects of various disturbances can never be completely excluded.
In particular, intrinsic turbulence effects can hardly be avoided in
chemical catalytic reactions or forest fires. One can also speculate that
atmospheric turbulence can play important role for the spreading of an
infectious disease by flying insects or birds. Effects of different kinds
of turbulent and laminar motion on the critical behavior were studied
e.g. in \cite{Satten}--\cite{AIK}.

In this paper we study effects of strongly anisotropic turbulent mixing
on the critical behavior of two paradigmatic models: the equilibrium
model {\it A}, which describes purely relaxational dynamics of a
nonconserved scalar order parameter (see e.g. \cite{Book3}), and the
Gribov model, which describes the nonequilibrium phase transition between
the absorbing and fluctuating states in a reaction-diffusion system
\cite{JT}. The velocity is modelled by the $d$-dimensional generalization
of the random shear flow introduced in \cite{AM}  within the context of
passive scalar advection.

In the Langevin formulation the models are defined by stochastic
differential equations for the order parameter
$\psi = \psi(t,{\bf x})$:
\begin{eqnarray}
\partial_{t} \psi = \lambda \left\{ (-\tau +
\partial^{2}) \psi - V(\psi) \right\} + \zeta  = 0,
\label{stoh}
\end{eqnarray}
where $\partial_{t}= \partial/ \partial t$, $\partial^{2}$ is the
Laplace operator, $\lambda>0$ is the kinematic (diffusion)
coefficient and $\tau \propto (T-T_{c})$ is the deviation of
the temperature (or its analog) from the critical value. The
nonlinearity has the form $V(\psi)=u \psi^{3}/3!$ for the model {\it A}
and $V(\psi)=g \psi^{2}/2$ for the Gribov process; $g$ and
$u>0$ being the coupling constants. The Gaussian random noise
$\zeta=\zeta(t,{\bf x})$ with zero mean is specified by the
pair correlation function:
\begin{eqnarray}
\langle \zeta (t,{\bf x})\zeta (t',{\bf x'}) \rangle =
2 \lambda  \delta(t-t')\delta^{(d)}({\bf x}-{\bf x}')
\label{kor1}
\end{eqnarray}
for the model {\it A} and
\begin{equation}
\langle \zeta (t,{\bf x})\zeta (t',{\bf x'}) \rangle = g\lambda\,
\psi(t,{\bf x})\,   \delta(t-t')\delta^{(d)}({\bf x}-{\bf x}')
\label{shum}
\end{equation}
for the Gribov process. The factor $\psi$ in front of the correlator
(\ref{stoh}) guarantees that in the
absorbing state the fluctuations cease entirely, while the factor
$2\lambda$ in (\ref{kor1}) ensures the correspondence to the static
$\psi^{4}$ model.

Coupling with the velocity field
${\bf v}=\{ v_{i}(t,{\bf x})\}$ is introduced by the replacement
$ \partial_{t} \to \nabla_{t} = \partial_{t} + v_{i} \partial_{i}$,
where $\nabla_{t}$ is the Lagrangian (Galilean covariant) derivative.

Let ${\bf n}$ be a unit constant vector that determines distinguished
direction
(``direction of the flow''). Then any vector can be decomposed into the
components perpendicular and parallel to the flow, for example,
${\bf x} = {\bf x}_{\bot} + {\bf n} x_{\parallel}$ with
${\bf x}_{\bot} \cdot {\bf n} =0$.
The velocity field will be taken in the form
$ {\bf v} =  {\bf n} v(t, {\bf x}_{\bot})$,
where  $v(t, {\bf x}_{\bot})$ is a scalar function
independent of $x_{\parallel}$. Then the incompressibility condition
$\partial_{i} v_{i} = \partial_{\parallel} v(t, {\bf x}_{\bot}) = 0$
is automatically satisfied.

For $v(t, {\bf x}_{\bot})$ we assume a Gaussian
distribution with zero mean and the pair correlation function of the form:
\begin{eqnarray}
\langle v(t,{\bf x}_{\bot}) v(t',{\bf x}_{\bot}') \rangle = D\,
\delta(t-t') \int \frac{d {\bf k}_{\bot}}{(2\pi)^{d-1}} \, \exp
\left\{ {\rm i} {\bf k}_{\bot}\cdot ({\bf x}_{\bot}-{\bf x}'_{\bot}) \right\}
\frac{1}{k_{\bot}^{d-1+\xi}}.
\label{veloc1}
\end{eqnarray}
Here $k_{\bot}=|{\bf k}_{\bot}|$,  $D>0$ is a constant amplitude factor
and $\xi$ an arbitrary exponent, which (along with the conventional
$\varepsilon=4-d$) will play the part of a formal RG expansion parameter.
The IR regularization
is provided by the cutoff $k_{\bot}>m$ in (\ref{veloc1}).
The natural interval for the exponent is $0< \xi <2 $ with
the most realistic Kolmogorov value $\xi=4/3$; the ``Batchelor limit''
$\xi\to2$ corresponds to smooth velocity.

According to the general theorem (see e.g. \cite{Book3}), stochastic problems
described above can be reformulated as field theoretic models of extended
set of fields $\Phi = \{\psi,\psi^{\dag},{\bf v}\}$. The role of the
coupling constants is played by the parameters
\begin{equation}
u, g^{2}  \sim \Lambda^{4-d}, \quad w = D/\lambda \sim \Lambda^{\xi},
\label{char}
\end{equation}
where $\Lambda$ is some typical ultraviolet (UV) momentum scale.
From (\ref{char}) it follows that the both models become logarithmic
(all the coupling constants are simultaneously dimensionless) at
$d=4$ and $\xi=0$. Thus the UV divergences in the Green functions manifest
themselves as poles in $\varepsilon = 4-d$, $\xi$ and, in general, their
linear
combinations. The divergences can be removed by the standard renormalization
procedure. In order to ensure {\it multiplicative} renormalizability
of the model, it is
necessary to split the Laplacian in (\ref{stoh}) into the parallel and
perpendicular parts $\partial^{2} \to \partial^{2}_{\bot}
+ f \partial^{2}_{\parallel}$ by introducing a new parameter
$f>0$ (in the anisotropic case, these two terms are renormalized
in a different way). Then the corresponding RG equations are derived;
their IR attractive fixed points determine the IR asymptotic scaling regimes
of the models. Detailed analysis of the model {\it A} can be found in
\cite{Alexa} and for the Gribov model it will be given elsewhere.
Below we only
present the results of the explicit one-loop calculation (leading order in
$\xi$ and $\varepsilon$), which appear similar for the both models.
There are four fixed points:

1) Gaussian (free) fixed point: $g_{*}=u_{*}=w_{*}=0$, IR attractive
for $\varepsilon<0$, $\xi<0$.

2) $g_{*}=u_{*}=0$ (exact result to all orders), $w_{*} \sim \xi$;
IR attractive for $\xi>0$ and  $\xi>2\varepsilon$. In this regime, the
nonlinearity $V(\psi)$ in the stochastic equation (\ref{stoh})
becomes irrelevant, and we arrive at the model of a linear
convection-diffusion equation for a passive scalar field $\psi$.

3) $w_{*}=0$ (exact), $g_{*}^{2}\sim u_{*} \sim\varepsilon$; IR attractive
for $\varepsilon>0$, $\xi<0$ for the model {\it A}
and $\varepsilon>0$, $\xi<\varepsilon/12$ for the Gribov model.
In this regime, effects of the turbulent mixing
are irrelevant, the isotropy violated by the velocity
ensemble is restored and the leading terms of the IR behavior coincide
exactly with those of the standard models (\ref{stoh})--(\ref{shum}).

4) The most interesting fixed point with nontrivial positive values of
$g_{*}$, $u_{*}$ and $w_{*}$. It is IR attractive for $\varepsilon>0$,
$\xi>0$ for the model {\it A} and $\varepsilon>0$, $\xi>\varepsilon/12$
for the Gribov model.
It corresponds to new nontrivial IR scaling regimes, in which both
nonlinearities in the stochastic equations and the turbulent mixing are
important; the corresponding critical dimensions depend essentially on
the both RG expansion parameters $\varepsilon$ and $\xi$ and are calculated
as double series in these parameters. This behavior reveals strong anisotropy
and belongs to new, completely nonequilibrium, universality classes.
The realistic values $\xi=4/3$ and $d=2$ or 3 belong to these classes.

The regions of IR stability for these fixed points are shown in Fig.~1.
For comparison, the boundaries for the isotropic stirring \cite{AIK}
are given. In the one-loop approximation, all the boundaries
are given by straight lines; there are neither gaps nor overlaps
between the different regions. The latter fact is valid to all orders
of the RG expansion, although the boundaries between the regions 3 and 4
(determined by the nonlinearity $V(\psi)$) and 2 and 4
(determined by the velocity statistics) will become curved.

Let us illustrate the consequences of the RG analysis for the spreading
of a cloud of the ``agent''  in the turbulent environment. The mean-square
displacement $R_{i}(t)$ in the $i$-th direction at time $t>0$ of the
agent's particle, which started from the origin ${\bf x'} = 0$ at time
$t'=0$, is given by the relation
\begin{eqnarray}
R^{2}_{i}(t) = \int d{\bf x}\ x^{2}_{i}\, G(t,{\bf x}), \quad
G(t,{\bf x}) = \langle \psi (t,{\bf x}) \psi^{\dag} (0,{\bf 0}) \rangle,
\quad x=|{\bf x}|.
\label{Rad}
\end{eqnarray}
The linear response function $G(t,{\bf x})$ has the following
asymptotic representation
\begin{eqnarray}
G(t,{\bf x}) =
x_{\bot}^{-\Delta_{\psi}-\Delta_{\psi^{\dag}}} F \left(
x_{\bot}t^{-1/\Delta_{\omega}},\, x_{\parallel}
t^{-\Delta_{\parallel}/\Delta_{\omega}}, \tau x_{\bot}^{\Delta_{\tau}}
\right),
\label{RAS}
\end{eqnarray}
where $x_{\bot}=|{\bf x}_{\bot}|$ and $F$ is the scaling function.
The set of critical dimensions $\Delta_{*}$ for the fields and parameters
is determined by the fixed point. In particular, for the Gribov model
at the fixed point 4 in the one-loop approximation one has
$\Delta_{\omega}=2-(2\varepsilon-\xi)/23$,
$\Delta_{\parallel}=1+ (12\xi-\varepsilon)/23$,
$\Delta_{\psi}=\Delta_{\psi^{\dag}}=d/2+ (14\xi-5\varepsilon)/46$,
$\Delta_{\tau}= 2+ 3(\xi-2\varepsilon)/23$.
Substituting (\ref{RAS}) into (\ref{Rad}) at $\tau=0$ (that is, directly at
criticality) gives the power laws
\begin{eqnarray}
R_{\bot}^{2}(t) \propto t^{\alpha_{\bot}}, \quad
R_{\parallel}^{2}(t) \propto t^{\alpha_{\parallel}},
\label{RA1}
\end{eqnarray}
where the exponents $\alpha_{\bot}$, $\alpha_{\parallel}$ are simply related
to the dimensions $\Delta_{*}$. For the fixed point 3 (linear passive
advection) one obtains exact results $\alpha_{\bot}=1$,
$\alpha_{\parallel}=1+\xi/2$. For the transverse direction this gives
the ordinary diffusion law $R_{\bot}(t) \propto t^{1/2}$, while in
the direction of the flow the spreading is accelerated and for $\xi=4/3$
takes on the form $R_{\parallel}(t) \propto t^{5/3}$, or equivalently
$d R_{\parallel}^{2} /dt \propto R_{\parallel}^{4/5}$. This ``4/5'' law
differs from the classical ``4/3'' Richardson's law for the isotropic
case. For the regime 4 the exponents in (\ref{RA1}) depend on the both
parameters $\xi$ and $\varepsilon$.

The authors thank the Organizers of the International Bogolyubov
Conference ``Problems of Theoretical and Mathematical Physics''
(Moscow--Dubna, 21--27 August 2009) for the possibility to present
the results of this work. The work was supported in part by the
RFFI grant No~08-02-00125a and the RNP grant No~2.1.1/1575. A.V.M.
was also supported by the Dynasty Foundation.

\newpage

\begin{figure}
\begin{center}
\includegraphics[width=14cm]{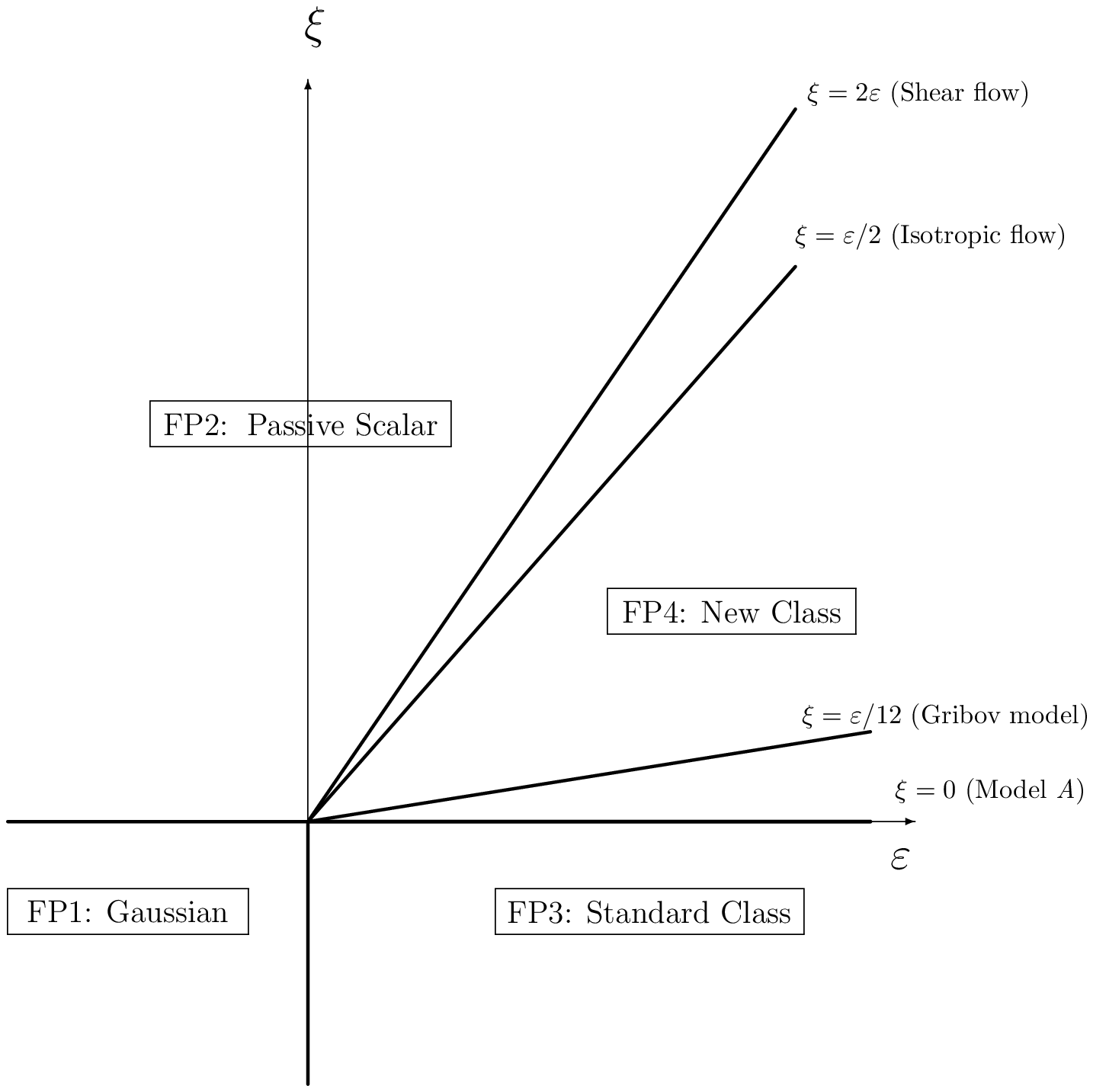}
\caption{Regions
of stability of the fixed points in various models.}
\end{center}
\end{figure}

\end{document}